\documentclass[fleqn,usenatbib,onecolumn]{mnras}

\usepackage{newtxtext,newtxmath}

\usepackage[T1]{fontenc}

\DeclareRobustCommand{\VAN}[3]{#2}
\let\VANthebibliography\thebibliography
\def\thebibliography{\DeclareRobustCommand{\VAN}[3]{##3}\VANthebibliography}

\usepackage{graphicx}	
\usepackage{amsmath}	

\title[An Optimal {\it In Situ} Multipole Algorithm for 3PCF]{An Optimal In-Situ Multipole Algorithm for the Isotropic Three-Point Correlation Function}

\author[Wenjie Ju et.al.]
{Wenjie Ju,
Longlong Feng\thanks{E-mail: flonglong@mail.sysu.edu.cn}, 
Zhiqi Huang, Xin Sun, Weishan Zhu
\\
$^{1}$School of Physics and Astronomy, Sun Yat-sen University, 2 Daxue Road, Tangjia, Zhuhai, 519082, China \\
}

\date{Accepted XXX. Received YYY; in original form ZZZ}

\pubyear{\the\year{}}

\begin{document}
\label{firstpage}
\pagerange{\pageref{firstpage}--\pageref{lastpage}}
\maketitle

\begin{abstract}

We present an optimised multipole algorithm for computing the three-point correlation function (3PCF), tailored for application to large-scale cosmological datasets. The algorithm builds on an {\it in situ} interpretation of correlation functions, wherein spatial displacements are implemented via translation window functions. In Fourier space, these translations correspond to plane waves, whose decomposition into spherical harmonics naturally leads to a multipole expansion framework for the 3PCF. To accelerate computation, we incorporate density field reconstruction within the framework of multiresolution analysis, enabling efficient summation using either grid-based or particle-based schemes. In addition to the shared computational cost of reconstructing the multipole-decomposed density fields—scaling as $\mathcal{O}(L^2_{\text{trun}} N_g \log N_g)$ (where $N_g$ is the number of grids and $L_{\text{trun}}$ is the truncation order of the multipole expansion)—the final summation step achieves a complexity of $\mathcal{O}(D^6_{\text{sup}} N_g)$ for the grid-based approach and $\mathcal{O}(D^3_{\text{sup}} N_p)$ for the particle-based scheme (where $D_{\text{sup}}$ is the support of the basis function and $N_p$ is the number of particles). The proposed {\it in situ} multipole algorithm is fully GPU-accelerated and implemented in the open-source {\tt Hermes} toolkit for cosmic statistics. This development enables fast, scalable higher-order clustering analyses for large-volume datasets from current and upcoming cosmological surveys such as Euclid, DESI, LSST, and CSST. 

\end{abstract}




\section{Introduction}

The widely accepted paradigm of cosmic structure formation is a hierarchical clustering model dominated by the cold dark matter driven by gravitational instability (e.g., \citealt{Efstathiou1990, Ostriker1995}). When validating structure formation models against large-scale galaxy surveys, the primary benchmarks are the two-point correlation function (2PCF) and its Fourier counterpart, the power spectrum, which serve as the fundamental clustering statistics for constraining both the cosmological parameters and the galaxy formation models (e.g., \citealt{Hamilton1988, Yang2003, cole2005, Eisenstein2005, Li2006, Tegmark2006, zehavi2011,Blake2011, Beutler2011, Percival2011, Anderson2014, Hildebrandt2016, Alam2017, Wechsler2018, Ivanov2020, Alam2021PhRvD}). However, as inhomogeneities grow and enter the nonlinear regime, gravitational collapse generates significant non-Gaussianities in the large-scale matter distribution across a wide range of scales. These non-Gaussian features—especially prominent at mildly and strongly nonlinear small scales due to mode-mode coupling—amplify both the variance and covariance of clustering measurements. Consequently, second-order statistics (e.g., 2PCF and the power spectrum) become insufficient to capture the full Fisher information content, posing a substantial challenge for constraining cosmological models using observational data \citep{Scoccimarro1999, Cooray2002, Rimes2005, Neyrinck2006, Nishimichi2016}.

To address this limitation, it is common to move beyond second-order statistics by employing polyspectra or higher-order correlation functions \citep{Peebles1980, Bernardeau2002}. In particular, the three-point correlation function (3PCF) and its Fourier-space counterpart—the bispectrum—offer enhanced sensitivity to both gravitational evolution and galaxy bias. On small scales, these higher-order statistics enable stringent tests of the cold dark matter paradigm and its alternatives (e.g., \citealt{Jing1998, Gaztanaga1994, Frieman1994, Scoccimarro2001, Jing2004, Kayo2004, PanSzapudi2005b, Nichol2006, GilMar2015, GilMar2017, Guo2015, Slepian2017b, Slepian2017a, Pearson2018, Slepian2018, Veropalumbo2021, Sugiyama2023}). On large scales, they are sensitive to primordial non-Gaussianity and can provide insights into the physics of the early universe (e.g., \citealt{Komatsu2001, Fergusson2009, chen2010, Desjacques2010, Scoccimarro2012, Biagetti2019, Meerburg2019, Achucarro2022} and references therein).

Moreover, the third- and higher-order statistics are powerful tools for probing and constraining galaxy bias models (e.g. \citealt{Fry1993, Desjacques2018} and references therein), and can compensate for the information loss in second-order statistics due to non-Gaussianity, significantly enhancing the precision of cosmological parameter estimation (e.g. \citealt{Gagrani2017, Agarwal2021, Alam2021_DESI, Gualdi2021, Samushia2021, Novell-Masot2023}). For example, joint analyses combining the power spectrum with the bispectrum and the trispectrum have been shown to yield tighter constraints on key cosmological parameters \citep{Gualdi2021}.

Technically, computing the N-point correlation function involves enumerating all possible N-tuples of galaxies, resulting in a brute-force computational complexity that scales as $\mathcal{O}(N_{p}^N)$ for a sample of $N_{p}$ galaxies. Therefore, early measurements of the 3PCF were limited to relatively small data sets containing only $\mathcal{O}(10^3)$ galaxies \citep{Groth_Peebles1977, Jing_Zhang1989}. However, the advent of next-generation redshift surveys—such as Euclid \citep{Laureijs2011}, the Dark Energy Spectroscopic Instrument (DESI) \citep{DESI2016}, CSST \citep{Zhan2011}, the Nancy Grace Roman Space Telescope \citep{Akeson2019}, and the Vera C. Rubin Observatory \citep{Ivezic2019}—which aim to map tens of millions to billions of galaxies, renders such brute-force approaches computationally prohibitive. This imposes a significant barrier to fully exploiting higher-order clustering statistics across wide ranges of configuration space and encoded physical information. 

To overcome this bottleneck, a variety of efficient algorithms have been developed to accelerate the evaluation of N-point correlation functions (NPCF). A common strategy involves organising the data into hierarchical spatial structures, which enable rapid neighbour searches and significantly reduce the number of necessary computations. Among these, the kd-tree has become a widely adopted framework for scalable N-point analyses in large galaxy catalogues \citep{Moore2001, Zhang2005,March2012,Sabiu2016,SabiuEtal2019}. Other strategies include pixelization schemes \citep{Gaztaga2009}, multipole decomposition \citep{Szapudi2004,Slepian2015,Slepian2018,PhilcoxEtal2022}, and Fourier transformation \citep{Slepian2016,Portillo2018,BrownEtal2022,PhilcoxSlepian2022,Sunseri2023}.

More recently, a {\it in situ} interpretation of the 2PCF was proposed by \citet{Yue2024}. In this framework, the correlation between two points separated by a fixed distance is reformulated as the correlation between the original density field and its translated counterpart. This translation is implemented via a convolution with a window function, which can be generalised to simultaneously incorporate multiple operations such as pair-counting, binning, and smoothing. This perspective not only provides a conceptually unified framework for understanding 2PCF, but also enables the definition of a broader class of generalised 2PCFs, each governed by a particular choice of window function.

When combined with multiresolution field decomposition, this {\it in situ} formulation yields a highly efficient algorithm for computing both standard and generalised 2PCFs. This methodology was initially implemented in the {\tt MRACS} (Multi-Resolution Analysis for Cosmic Statistics) framework \citep{Feng2007}, and has since evolved into the {\tt Hermes} toolkit (HypER-speed MultirEsolution cosmic statistics)—an open-source, GPU-accelerated Python package for cosmic statistical analysis (Ju et al., in preparation). Hereafter, we refer to its Python implementation as {\tt PyHermes}. Notably, this framework supports flexible, non-sharp binning schemes, enabling higher signal-to-noise measurements and improved precision.

Extending this idea, \citet{Yue2024} also introduced a fast algorithm to calculate the 3PCF. Unlike traditional approaches that rely on binning schemes tied to parameterisation of given triangle configurations, this method allows for assigning arbitrary window functions (e.g., spherical tophats or Gaussians) to each vertex of the triangle. Triplet counting is then performed efficiently using the count-in-cell techniques provided by the {\tt MRACS} framework and is subsequently entered into the optimal NPCF estimator of \citet{SzapudiSzalay1998}. However, since this method is essentially a Monte Carlo experiment, its convergence can vary significantly depending on the geometry of the triangle. Ensuring stable and accurate estimates thus requires case-by-case numerical tests. Accordingly, an open question remains: can a fast 3PCF algorithm be developed that is robust to variations in convergence rate without requiring case-specific adjustments? 

In this paper, we propose an optimal algorithm for estimating the isotropic 3PCF without explicit triplet counting. Our approach builds on the {\tt MRACS} framework, which decomposes the density field in terms of a complete set of compactly supported basis functions. The algorithm hinges on two core principles: (1) the ergodic assumption, which allows ensemble averages to be replaced by spatial averages, and (2) the orthogonality of the basis functions, which enables fast pair-counting through inner products between filtered coefficient vectors. Although the 2PCF estimator benefits from the direct integrability of the product of two basis functions, extending this to the 3PCF poses a key challenge: the product of three basis functions is generally not integrable. To overcome this, we adopt the concept of connection coefficients to approximate nonlinear operations in terms of linear combinations of decomposition coefficients. As we demonstrate in the following sections, this formulation leads to a novel, near-optimal algorithm for computing the multipole moments of the 3PCF. The resulting algorithm scales as $\mathcal{O}(D_{\text{sup}}^6 N_g)$, where $D_{\text{sup}}$ denotes the support of the basis functions, and $N_g$ is the number of spatial grids. Importantly, the algorithm’s complexity is independent of the number of particles, making it particularly suited for large-scale cosmological datasets.

The paper is organised as follows. In Section 2, we introduce several key concepts underlying our fast algorithm, including spatial translation operation and the {\it in situ} perspective of correlation functions, the spherical harmonic expansion of the window function, and the continuous multiresolution reconstruction of density fields from discrete point distributions. Based on these mathematical foundations, we develop an optimal multipole algorithm to measure the 3PCF. In Section 3, we apply the {\it in situ} multipole algorithm to cosmological simulation data and present benchmark performance tests. Finally, Section 4 summarises the results and makes concluding remarks.   

\section{An {\it In-Situ} Approach: Mathematics and Algorithm}

\subsection{2PCF and auto-correlation of scaling function basis}
\label{sec:maths} 

In cosmic statistics, a galaxy catalogue is currently taken to be a Poisson sampling of the underlying matter distribution. Mathematically, it corresponds to a spatial point process modelled by a number density field.
\begin{equation}\label{eq:Nsum-Dirac}
n({\bf x}) = \sum_{i=1}^{N_p} w_i \delta_D^3({\bf x} - {\bf x}_i),
\end{equation}
where $\{{\bf x}_i, w_i, i=1... N_p\}$ denotes the set of 3-dimensional coordinates and associated weights for $N_p$ discrete points. The 2PCF $\xi({\bf r})$ characterises the joint probability of finding a pair of points separated by a spatial displacement ${\bf r}$, within infinitesimal volume elements $d\tau_1$ and $d\tau_2$, such that 
\begin{equation}\label{eq:ex-situ-2PCF}
\langle dP \rangle = \langle n({\bf x}) n({\bf x} + {\bf r}) \rangle d\tau_1 d\tau_2 = \bar{n}^2 d\tau_1 d\tau_2 \left[1 + \xi({\bf r})\right],
\end{equation}
where $\bar{n}$ is the mean number density. Equation~\eqref{eq:ex-situ-2PCF} expresses the conventional {\it ex situ} definition of the 2PCF, in which $\xi({\bf r})$ quantifies the excess probability of finding a pair of points at a given separation ${\bf r}$, relative to a Poisson (random) distribution.

In contrast, the {\it in situ} perspective reinterprets the 2PCF as the correlation between the original density field $n({\bf x})$ and its translation copy, 
\begin{equation}
\tilde{n}_{\bf r}({\bf x})=n({\bf x}+{\bf r})=\hat{\mathscr{W}}({\bf r})n({\bf x}), \quad \hat{\mathscr{W}}({\bf r})=e^{i\hat{\bf p}\cdot{\bf r}}
\end{equation}
where $\hat{\mathscr{W}}({\bf r})$ produces the spatial translation generated by the momentum operator $\hat{\bf p}=-i{\boldsymbol{\nabla}}_{\bf x}$. Alternatively, translation can also be achieved through a convolution using the window function $W_{\bf r}$, 
\begin{equation}
\tilde{n}_{\bf r}({\bf x}) = \hat{W}_{\bf r} ({\bf x})\circ n({\bf x}) \equiv\int \hat{W}_{\bf r}({\bf x}-{\bf x}')n({\bf x}')d^3{\bf x}', \quad \hat{W}_{\bf r}({\bf x})=\delta^3_D({\bf x}+{\bf r})
\end{equation}

Suppose that the point process $n({\bf x})$ is modulated by a underlying random field $\delta({\bf x})$ such that $n({\bf x})=\bar{n}[1+\delta({\bf x})]$, the 2PCF is related to its Fourier counterpart-the power spectrum,
\begin{equation}
\xi({\bf r}) = \langle \delta({\bf x})\tilde{\delta}_{\bf r}({\bf x})\rangle=\int P({\bf k})\hat{W}_{\bf r}({\bf k})d^3{\bf k},
\end{equation}
where $P({\bf k})=|\delta_{\bf k}|^2$ is the power spectrum of $\delta({\bf x})$ under the assumption of statistical homogeneity, and $\hat{W}_{\bf r}(\bf k)$ is the Fourier transform of the window function, given by
\begin{equation}\label{eq:window-sphshell}
    \hat{W}_{\bf r}(\bf k)=\text{e}^{i{\bf k}\cdot{\bf r}}.
\end{equation}
Furthermore, if $n({\bf x})$ is also statistically isotropic, $\xi({\bf r})$ depends only on the scalar distance $r=|{\bf r}|$. In this case, the angular average over the spherical shell of radius $r$ yields,
\begin{equation}
 \langle\hat{W}_{\bf r}({\bf k}) \rangle_{\text{sph-shell}} = \langle \text{e}^{i{\bf k}\cdot{\bf r}}\rangle_{\text{sph-shell}} =\frac{1}{4\pi}\int_0^{2\pi}d\phi\int_0^{\pi} \text{e}^{ikr\cos\theta}\sin\theta d\theta   =\frac{\sin kr}{kr},  
\end{equation}
which recovers the current relation of the 2PCF and the power spectrum. Correspondingly, we can find the real-space representation of the spherically averaged window function,  
\begin{equation}\label{eq:window-sphshell-real}
  \langle{W}_{\bf r}({\bf x}) \rangle_{\text{sph-shell}}=\frac{\delta(|{\bf x}|-r)}{4\pi r^2}
\end{equation}
which describes the volume density of an infinitesimally thin spherical shell of radius $r$.

Building on the {\it in situ} interpretation of the 2PCF, we can realise a fast algorithm for pair counting as implemented in the {\tt Hermes} framework \citep{Feng2007, Yue2024}. The key idea is to represent both the original density field and its filtered (translated) counterpart using a set of compactly supported scaling basis functions $\phi_{j{\bf l}}$, such that      
\begin{equation}
    n({\bf x})=\sum_{\bf l}\epsilon_{j{\bf l}} ({\bf x})\phi_{{j{\bf l}}}({\bf x}), \quad 
    \tilde{n}_{\bf r}({\bf x})=\sum_{\bf l}\tilde\epsilon_{j{\bf l}}({\bf r})\phi_{{j{\bf l}}}({\bf x}).
\end{equation}
where $\epsilon_{j\mathbf{l}}$ and $\tilde{\epsilon}_{j\mathbf{l}}(\mathbf{r})$ are the scaling function coefficients (SFCs) of the original and translated fields, respectively.

Under the ergodic assumption, the ensemble average can be equivalently converted to averaging over a sampling volume, thus the ensemble-average pair-counting can be given by
\begin{equation}\label{eq:pair-counting}
    DD = \frac{1}{V}\int n({\bf x})\tilde{n}_{\bf r}({\bf x}) d^3{\bf x} = \sum_{{\bf l}}\sum_{{\bf l'}} \epsilon_{j{\bf l}} \tilde\epsilon_{j{\bf l'}}({\bf r})\frac{1}{V}\int \phi_{j{\bf l}}({\bf x})\phi_{j{\bf l'}}({\bf x})d^3{\bf x} = \sum_{{\bf l}}\epsilon_{j{\bf l}} \tilde\epsilon_{j{\bf l}}({\bf r}),
\end{equation}
where the orthogonality of the basis functions has been used. Eq.(\ref{eq:pair-counting}) implies that the pair-counting can be simply evaluated by a scalar product of two vectors $\boldsymbol{\epsilon}=\{\epsilon_{j{\bf l}}\}$ and $\boldsymbol{\tilde\epsilon}=\{\tilde\epsilon_{j{\bf l}}\}$. 

For the translation field $\tilde{n}_{\bf r}$, it is easy to show that the SFC $\tilde\epsilon_{\bf jl}({\bf r})$ of the translation field would be
\begin{equation}\label{eq:shift-scaling_coeff}
    \tilde\epsilon_{j{\bf l}}({\bf r}) = \frac{1}{V}\int d{\bf x}n({\bf x+r})\phi_{j{\bf l}}({\bf x}) = \sum_{\Delta {\bf l}}{\epsilon}_{j{\bf l}+\Delta {\bf l}}{\Phi}_{\Delta{\bf l}}({\bf r}),
\end{equation}
where $\Phi_{\Delta {\bf l}}({\bf r}) \equiv \Phi({\bf r}-\Delta{\bf l})$, $\Phi({\bf r})$ is the auto-correlation of the scaling function basis
\begin{equation}\label{eq:cf-basis}
    \Phi({\bf r})=\frac{1}{V}\int d{\bf s}\phi({\bf s})\phi({\bf s+r}).
\end{equation}
Applying the discrete Fourier transformation (DFT) leads to an expansion in the wavenumber space,   
\begin{equation}\label{eq:trans-field-FT}
\begin{aligned}
\tilde\epsilon_{j \mathbf{l}}(\mathbf{r}) &= \sum_{\Delta \mathbf{l}} \epsilon_{j, \mathbf{l}+\Delta \mathbf{l}} \hat\phi(\mathbf{r}-\Delta \mathbf{l}) 
= \sum_{\Delta \mathbf{l}} \sum_{\mathbf{m}} \hat{\epsilon}_{j \mathbf{m}} e^{i{\mathbf k}_{\mathbf m}\cdot(\mathbf{l}+\Delta \mathbf{l})} \sum_{\mathbf{n}} \hat\Phi_{\mathbf{n}} e^{i{\mathbf k}_{\mathbf n}\cdot(\mathbf{r}-\Delta \mathbf{l})}  \\
&= \sum_{\mathbf{m}} \hat{\epsilon}_{j \mathbf{m}}e^{i\mathbf{k}_{\mathbf m}\cdot \mathbf{l}}\Bigl[\sum_{\mathbf{n}}  \hat\Phi_{\mathbf{n}}  e^{i{\mathbf k}_{\mathbf n}\cdot \mathbf{r}} \sum_{\Delta \mathbf{l}} e^{i({\mathbf k}_{\bf m}-{\mathbf k}_{\mathbf n})\cdot \Delta \mathbf{l}}\Bigr] 
= \sum_{\mathbf{m}} \hat{\epsilon}_{j \mathbf{m}} e^{i{\mathbf k}_{\bf m} \cdot \mathbf{l}} \Bigl[\sum_{\mathbf{n}}  \hat\Phi_{\mathbf{n}}  e^{i{\mathbf k}_{\mathbf n} \cdot \mathbf{r}} \delta_{\mathbf{m}, \mathbf{n}}\Bigr] \\
&= \sum_{\mathbf{m}} \hat{\epsilon}_{j \mathbf{m}} \hat\Phi_{\mathbf{m}} e^{i \mathbf{k_m} \cdot \mathbf{r}} e^{i \mathbf{k_m} \cdot \mathbf{l}}, \\
\end{aligned}
\end{equation}
where the DFT wavenumbers are ${\bf k}_{\bf m}= \frac{2\pi}{L}{\bf m}$.
Therefore, using the above equation Eq.(\ref{eq:trans-field-FT}), the pair-counting Eq.(\ref{eq:pair-counting}) can be further given by the inverse DFT summation,
\begin{equation}
  DD = \sum_{{\bf m},{\bf m}'}\sum_{\bf l}\hat{\epsilon}_{j{\bf m}'}^*\hat\epsilon_{j{\bf m}}\hat{\Phi}_{\bf m}\text{e}^{i{\bf k}_{\bf m}\cdot{\bf r}}\text{e}^{i({\bf k}_{{\bf m}}-{\bf k}_{{\bf m}'})\cdot{\bf l}}
  =\sum_{\bf m} |\hat\epsilon_{j{\bf m}}|^2\hat{\Phi}_{\bf m}\text{e}^{i{\bf k}_{\bf m}\cdot{\bf r}}.
\end{equation}
If the system possesses some sort of spatial symmetries, we can perform the associated spatial average correspondingly,    
\begin{equation}\label{eq:DD-multipole}
    \langle DD \rangle =\sum_{\bf m} |\hat\epsilon_{j{\bf m}}|^2\hat{\Phi}_{\bf m} \langle \text{e}^{i{\bf k}_{\bf m}\cdot{\bf r}} \rangle.
\end{equation}
Assuming the axial symmetry that is commonly assumed in the redshift space and adopting spherical coordinates $\{r,\theta_{\bf r}, \phi_{\bf r}\}$, the spatial average can be performed by integrating along the azimuthal direction,
\begin{equation}
\langle \text{e}^{i{\bf k}_{\bf m}\cdot{\bf r}}\rangle_{\phi_{\bf r}} 
 = \frac{1}{2\pi}\int_0^{2\pi} \text{e}^{i{\bf k}_{\bf m}\cdot{\bf r}} d{\phi_{\bf r}}
=\displaystyle{\sum_{\ell=0}^{\infty}}\hat{W}_r^{\ell}({\bf k}_{\bf m})P_\ell(\cos\theta_{\bf r}),
\end{equation}  
with the multipole components of the spherical-shell filter, defined by 
\begin{equation}\label{eq:DD-windowfunction}
   \hat{W}_r^{\ell}({\bf k}_{\bf m})= (2\ell+1)i^{\ell} j_\ell(k_{\bf m}r)P_\ell(\cos{\theta}_{\bf m}),
\end{equation}
in which we have used the spherical harmonics expansion of plane waves,
\begin{equation}\label{eq:plane-sph-harmonics}
    e^{i \mathbf{k} \cdot \mathbf{r}}=4 \pi \sum_{\ell=0}^{\infty} \sum_{m=-\ell}^l i^\ell j_\ell(k r) Y_\ell^m\left(\Omega_k\right)^* Y_\ell^m\left(\Omega_r\right) .
\end{equation}
Eq.(\ref{eq:DD-multipole})-(\ref{eq:DD-windowfunction}) give a multipole algorithm for 2PCF in redshift space. In Fourier space, multipole algorithms for the anisotropic power spectrum have been developed in previous works (e.g., \citealt{Bianchi2015, Scoccimarro2015, Hand2017}). In the case of a spherically symmetric system, only the monopole component survives $\hat{W}_r^0({\bf k}_{\bf m})=j_0(k_{\bf m}r)=\frac{\sin(k_{\bf m}r)}{k_{\bf m}r}$, recovering again the window function Eq.(\ref{eq:window-sphshell}) for an infinitely thin spherical shell of radius $r$.   

\subsection{3PCF and connection coefficients of basis functions}
\label{subsec:3PCF}

The {\it in situ} expression of 2PCF can also be extended to estimate the monopole moment in high-order correlation functions. We demonstrate a similar summation rule for the monopole moment in the three-point correlation function. In the meaning of the angular average, the monopole component can be estimated by pair counting in two neighbour shells $R_1$ and $R_2$ around a given point. Explicitly, we have
\begin{equation}\label{eq:3pcfddd}
DDD = \langle n({\bf x})\tilde{n}_{R_1}({\bf x})\tilde{n}_{R_2}({\bf x})\rangle = \frac{1}{V}\int n({\bf x})\tilde{n}_{R_1}({\bf x})\tilde{n}_{R_2}({\bf x}) d{\bf x} ,
\end{equation}
where $\tilde{n}_{R_i}({\bf x})$ is the pair count in the shell $R_i$, $\tilde{n}_{R_i}({\bf x}) = [W_{\text{shell}}({\bf x},R_i)\circ n]({\bf x}), i=1,2$. In terms of a set of bases $\{\phi_{j{\bf l}}\}$, the corresponding SFCs are $\{\epsilon_{j{\bf l}}\}$, $\{\tilde\epsilon^{R_1}_{j{\bf l}}\}$,$\{\tilde\epsilon^{R_2}_{j{\bf l}}\}$, such that
$\tilde{n}_{R_i}({\bf x}) = \sum_{j{\bf l}}\tilde\epsilon^{R_i}_{j{\bf l}}\phi_{j{\bf l}}$. 
Introducing the connection coefficients $\Gamma_{{\bf l}_1,{\bf l}_2}$ defined by 
\begin{equation}\label{eq:con-coeff}
  \Gamma_{{\bf l}_1,{\bf l}_2} = \frac{1}{V}\int d{\bf x}\phi({\bf x})\phi({\bf x-l}_1)\phi({\bf x-l}_2),
\end{equation}
the {\it in situ} twofold density $\tilde{n}_{R_1}({\bf x})\tilde{n}_{R_i}({\bf x})$ can be decomposed in terms of basis functions $\phi_{j{\bf l}}({\bf x})$ again, 
\begin{equation}\label{eq:2fold-monopole}
    n_{\text{2-fold}}^{R_1, R_2}({\bf x}) \equiv \tilde{n}_{R_1}({\bf x})\tilde{n}_{R_2}({\bf x})=\sum_{\bf l}\tilde\epsilon_{j,{\bf l}}(R_1,R_2)\phi_{j,{\bf l}}({\bf x}),
\end{equation}
where
\begin{equation}
   \tilde\epsilon_{j,{\bf l}}(R_1,R_2)=\frac{1}{V}\int \tilde{n}_{R_1}({\bf x})\tilde{n}_{R_2}({\bf x})\phi_{j{\bf l}}({\bf x})d^3{\bf x}
   =  \sum_{{\bf l}_1}\sum_{{\bf l}_2} \Gamma_{{\bf l}_1,{\bf l}_2}\tilde\epsilon^{R_1}_{j{\bf l+l}_1}  \tilde\epsilon^{R_2}_{j{\bf l+l}_2}.
\end{equation}
It is noted that the summations of ${\bf l}_1$ and ${\bf l}_2$ are taken over only a limited interval depending on the support domain of the basis functions. The computational complexity of summation is $\mathcal{O}(D_{\text{sup}}^{6} N_g)$, where $N_{g} = 2^{3j}$, $D_{\text{sup}}$ is the linear size of the support of the basis function. For the Daubechies wavelet with the genus $D_g=4$, $D_{\text{sup}}=D_g-1=3$, the complexity scales roughly as $\sim 10^3N_{g}$ in 3-dimensional space. 

Similarly, for the primary vertex ${\bf x}$, we also have $n({\bf x})= \sum_{j{\bf l}}\epsilon_{j{\bf l}}\phi_{j{\bf l}}({\bf x})$. 
Therefore,  
\begin{equation}\label{eq:monopole_tri_counting}
    DDD= \frac{1}{V}\int n({\bf x})n^{R_1,R_2}_{\text{2-fold}}({\bf x})d^3{\bf x} = \sum_{\bf l} \epsilon_{j{\bf l}} \tilde\epsilon_{j,{\bf l}}(R_1,R_2).
\end{equation}
Remarkably, as shown in Eq.(\ref{eq:monopole_tri_counting}), the triplet counting in the 3PCF estimation has been degraded and reformulated into a structure analogous to the pair counting used in the 2PCF estimation (Eq.\ref{eq:pair-counting}). Specifically, it is compactly expressed as a scalar product (dot) between two coefficient vectors, $\boldsymbol{\epsilon} = \{\epsilon_{j{\bf l}}\}$ and $\boldsymbol{\tilde{\epsilon}} = \{\tilde{\epsilon}_{j{\bf l}}\}$.

In the same spirit of the multipole decomposition of 2PCF, the above algorithm can be extended to estimate 3PCF for arbitrary triangle configurations. Generally, for triplet counting, the spherical radius $R_1$ and $R_2$ in Eq.(\ref{eq:3pcfddd}) are replaced by two translation vectors ${\bf r}_1$ and ${\bf r}_2$.  
\begin{equation}\label{eq:3PCF-spatial-average}
DDD  = \langle n({\bf x})\tilde{n}_{{\bf r}_1}({\bf x})\tilde{n}_{{\bf r}_2}({\bf x})\rangle = \frac{1}{V}\int n({\bf x}){n}({\bf x}+{\bf r}_1){n}({\bf x}+{\bf r}_2) d^3{\bf x}.
\end{equation}
In terms of the basis of scaling functions, DFT leads to
\begin{equation}\label{eq:DDD-3PCF-DFT}
\begin{aligned}
DDD  =&\sum_{\mathbf{l}} \epsilon_{j \mathbf{l}} \sum_{\mathbf{l}_1} {\epsilon}_{j \mathbf{l}+\mathbf{l}_1}^{{\bf r}_1} \sum_{\mathbf{l}_2} {\epsilon}_{j \mathbf{l}+\mathbf{l}_2}^{{\bf r}_2}\Gamma_{\mathbf{l}_1, \mathbf{l}_2}  
=\sum_{\mathbf{l}} \epsilon_{j \mathbf{l}} \sum_{\mathbf{l}_1} \sum_{\mathbf{m}} \hat{\epsilon}_{j \mathbf{m}} \hat\phi_{\mathbf{m}} \text{e}^{i \mathbf{k_m}\cdot (\mathbf{l}+\mathbf{l_1}+\mathbf{r_1})}\sum_{\mathbf{l}_2}\sum_{\mathbf{n}} \hat{\epsilon}_{j \mathbf{n}} \hat\phi_{\mathbf{n}} \text{e}^{i\mathbf{k_n}\cdot (\mathbf{l}+\mathbf{l_2}+\mathbf{r_2})}\Gamma_{\mathbf{l}_1, \mathbf{l}_2}   \\ 
=&\sum_{\mathbf{l}} \epsilon_{j \mathbf{l}} \sum_{\mathbf{l}_1,\mathbf{l}_2} \Gamma_{\mathbf{l}_1, \mathbf{l}_2} \sum_{\mathbf{m}} \hat{\epsilon}_{j \mathbf{m}} \hat\phi_{\mathbf{m}} \text{e}^{i\mathbf{k_m} \cdot(\mathbf{l}+\mathbf{l_1})}\sum_{\mathbf{n}} \hat{\epsilon}_{j \mathbf{n}} \hat\phi_{\mathbf{n}} \text{e}^{i\mathbf{k_n}\cdot (\mathbf{l}+\mathbf{l_2})}e^{i\mathbf{k_m}\cdot \mathbf{r_1}}e^{i\mathbf{k_n}\cdot \mathbf{r_2}}.  
\end{aligned}
\end{equation}
Since triangle configuration can be specified by the lengths of two sides and the angle between them, the 3PCF can be parameterised by three parameters $\{r_1=|{\bf r}_1|,r_1=|{\bf r}_1|, \cos\theta=\hat{\bf r}_1\cdot\hat{\bf r}_2\}$. By averaging the spatial orientations of triangles of a given configuration, we can obtain the following useful formula after some algebra,  
\begin{equation}\label{eq:double-e-expansion}
\langle \text{e}^{i{\bf k}_1 \cdot {\bf r}_1} e^{i{\bf k}_2 \cdot {\bf r}_2} \rangle = \sum_{L=0}^{\infty} (-1)^L (2L + 1) P_L (\hat{\bf k}_1 \cdot \hat{\bf k}_2) P_L (\hat{\bf r}_1 \cdot \hat{\bf r}_2) j_L(k_1 r_1) j_L(k_2 r_2)=\sum_{L=0}^{\infty}\Bigl(\sum_{M=-L}^{L}\hat{W}^{LM}_{r_1}({\bf k}_1)\hat{W}^{LM}_{r_2}({\bf k}_2)\Bigr) P_L (\hat{\bf r}_1 \cdot \hat{\bf r}_2),
\end{equation}
with the multipole window function of the spherical shell in wavenumber space defined by,
\begin{equation}\label{eq:3PCF-multipole-window}
    \hat{W}^{LM}_{r}({\bf k})= \sqrt{4\pi}i^{L}j_L(kr)Y_{L}^{M}(\hat{\bf k}).
\end{equation}
The proof of the above formula is given briefly in Appendix~\ref{sec:proofs}. Substituting Eq.(\ref{eq:double-e-expansion}) into Eq.(\ref{eq:DDD-3PCF-DFT}) yields the multipole decomposition of the triplet counting, 
\begin{equation}\label{eq:triplet-counting}
\langle DDD \rangle = \sum_{L=0}^{\infty}\zeta_L^{DDD}(r_1,r_2) P_L (\hat{\mathbf r}_1 \cdot \hat{\mathbf r}_2) 
\end{equation}
where the multipole components are given by
\begin{equation}\label{eq:3PCF-multipole-sum}
    \zeta_L^{DDD}(r_1,r_2) = \sum_{\mathbf{l}}\epsilon_{j \mathbf{l}}\tilde\epsilon_{j\mathbf{l}}^{L},(r_1,r_2)
\end{equation}
in which  
\begin{equation}\label{eq:SFC-3PCF}
    \tilde\epsilon_{j\mathbf{l}}^{L}(r_1,r_2) = \sum_{M=-L}^{L} \sum_{\mathbf{l}_1,\mathbf{l}_2} \Gamma_{\mathbf{l}_1, \mathbf{l}_2}\tilde{\epsilon}_{j \mathbf{l}+\mathbf{l}_1}^{LM}(r_1)  \tilde{\epsilon}_{j \mathbf{l}+\mathbf{l}_2}^{LM}(r_2) 
\end{equation}
with 
\begin{equation}\label{eq:SFC-3PCF-vertex}
    \tilde{\epsilon}_{j \mathbf{l}}^{LM}(r)
    =\sum_{\mathbf{m}}\hat{\epsilon}_{j \mathbf{m}}\hat\Phi_{\mathbf{m}}\hat{W}^{LM}_{\bf r}({\bf k}_{\bf m})e^{i\mathbf{k_{\bf m}}\cdot\mathbf{l}} 
\end{equation}
Eq.(\ref{eq:SFC-3PCF-vertex}) gives the convolved SFCs at the given harmonic mode $(L,M)$, which can be numerically evaluated using DFT.  
    
As shown in Eq.(\ref{eq:SFC-3PCF}), $\epsilon^L_{j{\bf l}}(r_1, r_2)$ characterises the $l$-th multipole moment of the rotationally averaged twofold density field for a given triangle configuration $(r_1, r_2)$. Similar to Eq.(\ref{eq:2fold-monopole}), we introduce the twofold density decomposed in the multipole space,    
\begin{equation}\label{eq:nL-2fold}
n^{L}_{\text{2-fold}}({\bf x},r_1,r_2) = \sum_{\bf l} \tilde{\epsilon}^L_{j{\bf l}}(r_1, r_2)\, \phi_{j{\bf l}}({\bf x}).
\end{equation}

As formulated in Eqs.~(\ref{eq:triplet-counting})–(\ref{eq:SFC-3PCF-vertex}), triplet counting can be performed directly on the reconstructed density field without requiring binning. In this case, each vertex in the triangle is treated as unfiltered, which means that no binning-window function is applied. However, in our algorithm, different binning window functions can be flexibly assigned to each vertex. These functions are then convolved with the original field to obtain the filtered field at the corresponding vertex. In practical implementation, this amounts to multiplying the multipole window function defined in Eq.~(\ref{eq:3PCF-multipole-window}) by the assigned binning window function at each vertex.   

In the special case where no binning window function is applied to the primary vertex, $n({\bf x})$ is thus given by Eq.~(\ref{eq:Nsum-Dirac}). Returning to the original integral expression for $DDD$ as given in Eq.(\ref{eq:monopole_tri_counting}), the multipole summation in Eq.~(\ref{eq:3PCF-multipole-sum}) can be equivalently rewritten as a sum over discrete objects in the data catalogue,
\begin{equation}\label{eq:DDD-sum-privertex}
\zeta_L^{DDD}(r_1, r_2) = \frac{1}{V} \int n({\bf x})\, n^L_{\text{2-fold}}({\bf x},r_1,r_2)\, d^3{\bf x} = \sum_{i=1}^{N_p}w_in^L_{\text{2-fold}}({\bf x_i},r_1,r_2) 
\end{equation}
where $n^L_{\text{2-fold}}({\bf x},r_1,r_2)$ is evaluated by Eq.~(\ref{eq:nL-2fold}) together with Eq.~(\ref{eq:SFC-3PCF})-(\ref{eq:SFC-3PCF-vertex}). The above equation offers an alternative form of spatial averaging. This formulation is particularly advantageous when the number of particles $N_p$ is significantly smaller than the total number of grids. In such cases, performing the summation over particles is computationally more efficient than evaluating over the full grid volume. Practically, the summation Eq.~(\ref{eq:DDD-sum-privertex}) can be further accelerated in the following way. Using Eq.~(\ref{eq:SFC-3PCF-vertex}), we can introduce formally a set of number densities in the spherical harmonic space labelled by $(L,M)$
\begin{equation}
 n^{LM}({\bf x},r)=\sum_{\bf l}\tilde{\epsilon}_{j{\bf l}}^{LM}(r)\phi_{j{\bf l}}({\bf x}) 
\end{equation}
According to Eq.~(\ref{eq:SFC-3PCF}), it is easy to see that $n^L_{\text{2-fold}}({\bf x},r_1,r_2)$ can be calculated alternatively by summing up twofold $n^{LM}({\bf x},r_1)n^{LM}({\bf x},r_2)$ in the spherical harmonic subspace in the given multipole order $L$, that is, 
\begin{equation}
n^L_{\text{2-fold}}({\bf x},r_1,r_2)=\sum_{M=-L}^{L} n^{LM}({\bf x},r_1) n^{LM}({\bf x},r_2)
\end{equation}
Thereby, rather than performing a summation over grid cells, we can reformulate the computation as a summation over particles, yielding the following expression:
\begin{equation}
\zeta_L^{DDD}(r_1, r_2) = \sum_{M=-L}^{L} \sum_{i=1}^{N_p} w_i n^{LM}({\bf x}_i, r_1) n^{LM}({\bf x}_i, r_2),
\end{equation}
where $n^{LM}({\bf x}_i, r)$ are evaluated only at the positions of particles ${\bf x}_i$. This formulation enables the construction of an optimal algorithm for estimating the Monte Carlo integral, achieving minimal computational cost. Aside from the comparable cost of performing FFT, the grid-based summation scales as $\mathcal{O}(L D_{\text{sup}}^6 N_g)$, while the particle-based summation scales as $\mathcal{O}(L D_{\text{sup}}^3 N_p)$. Thus, when the number of particles is less than the number of grids, $N_p \leq N_g$, the acceleration factor of the particle-based implementation is roughly $\sim D_{\text{sup}}^3 N_g / N_p \gg 1$.   

\section{Implementation and Numerical Test}
\label{NumericalTest}

\begin{figure*}
    \centering
    \includegraphics[width=\textwidth]{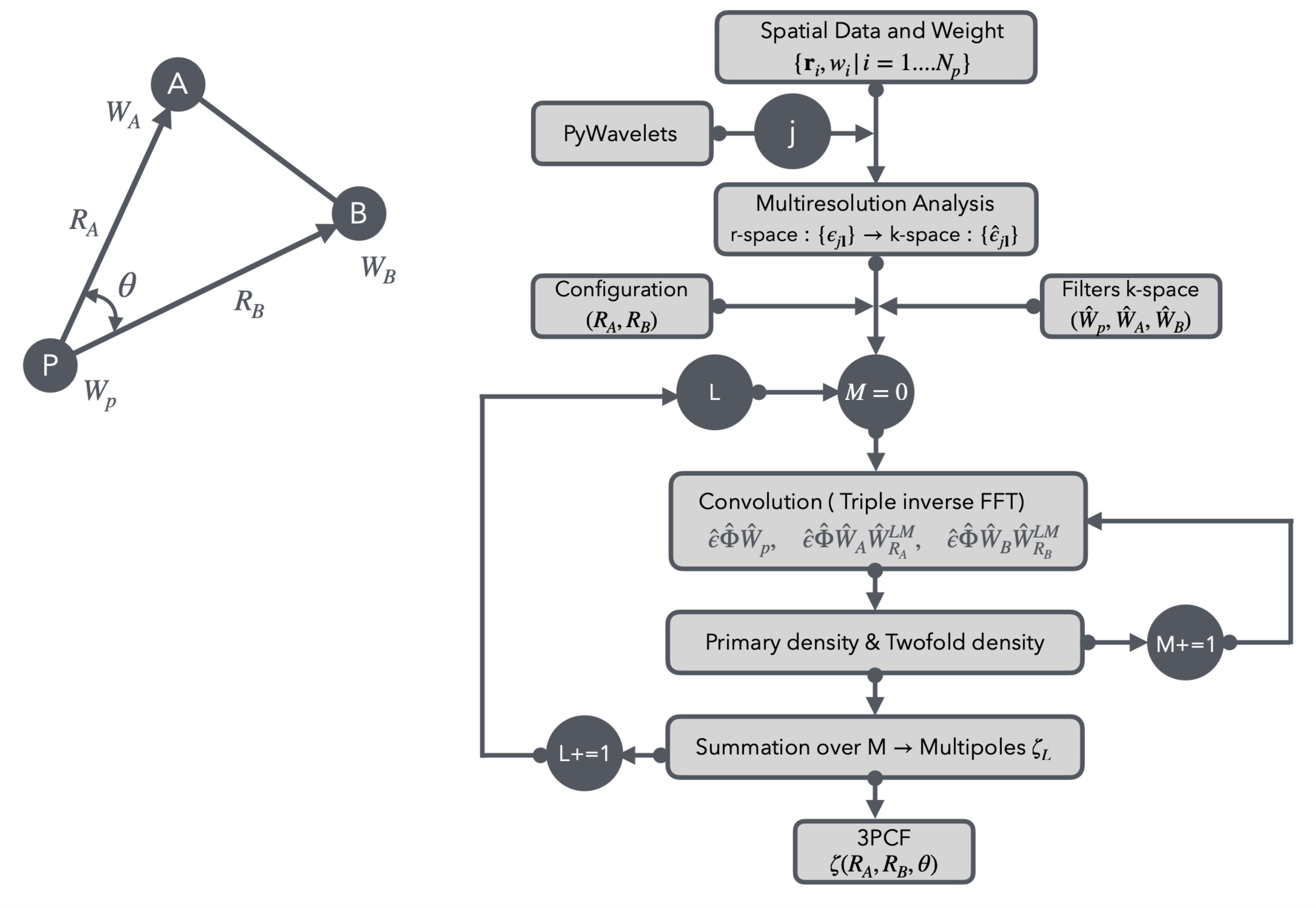}
    \vspace{-0.5cm}
    \caption{
    The flowchart illustrates the practical implementation of the {\it in situ} multipole method, which proceeds through the following steps: \\
    (1) Initialization: The input on the user-end interface includes the data catalogue, selecting basis functions from the wavelet library, specifying the binning window function $\Phi$ and the triangle configuration $\{R_A, R_B\}$. \\
    (2) Density field decomposition in multiresolution space: The discrete spatial data is decomposed in terms of the scaling function basis: $\{{\bf r}_i, w_i\} \rightarrow {\epsilon_{j{\bf l}}}$, and then transformed to the wavenumber k-space $\{\hat\epsilon_{j{\bf l}}\}$ via FFT. \\
	(3) Tabulating window functions in wavenumber space: We generate arrays for the window functions including the power spectrum of father scaling function $\hat\Phi$, the binning (smoothing) functions on the primary vertex, $\hat{W}_p$, and on the two secondary vertices, $\hat{W}_A$ and $\hat{W}_B$.\\
	(4) Generating rotation-averaged translation window functions in spherical harmonic space: For a given spherical harmonic mode $(L, M)$, we construct the rotation-averaged translation window functions $\hat{W}^{LM}_{R_A}$ and $\hat{W}^{LM}_{R_B}$ in $k$-space, and then combine them with the binning functions at the corresponding secondary vertices to yield composite kernels: $\hat{W}_A \hat{W}^{LM}_{R_A}$ and $\hat{W}_B \hat{W}^{LM}_{R_B}$, respectively.\\
	(5) Computing the SFCs in real space: The SFCs in k-space are convolved with the corresponding window functions generated in steps (3) and (4) via inverse FFT to produce three sets of spatial fluctuation coefficients (SFCs): ${\epsilon_{j{\bf l}}, \epsilon_{j{\bf l}}^{R_A}, \epsilon_{j{\bf l}}^{R_B}}$.\\
	(6) Constructing the twofold density to measure multipole moments: Looping over $M$ for a given $L$, we generate the twofold {\it in situ} density via either grid-based or particle-based summation, enabling the measurement of the corresponding $L$-mode multipole moment. \\
    (7) Measuring the multipole moments and the 3PCF: By looping over $L$ up to a specified truncation order, we compute the corresponding multipole moments, yielding a compressed representation of the 3PCF. By summing over the multipole modes, the full 3PCF can be reconstructed, enabling the investigation of its shape dependence.
    }
    \label{fig:3PCF-pipline}
\end{figure*}

We implement the multipole algorithm described in Section 2.2 in Python with GPU acceleration and integrate it into the {\tt PyHermes} toolkit. The overall workflow is illustrated in Fig.~\ref{fig:3PCF-pipline}. The user interface requires three key inputs: (1) a catalogue containing spatial coordinates and weights for each object, $\{{\bf r}_i, w_i \mid i = 1, \ldots, N\}$; (2) a set of basis functions $\{\Phi_{j{\bf l}}\}$; and (3) binning window functions ${W_p, W_A, W_B}$ associated with the three vertices of the given triangle configuration.

In this study, we used the MultiDark Planck 2 (MDPL2) simulation \footnote{\url{https://www.cosmosim.org}} to evaluate the precision and performance of our multipole algorithm for 3PCF. The MDPL2 simulation is part of the MultiDark project \citep{prada2012halo, riebe2013multidark}, designed to provide high-resolution $N$-body realisations of the large-scale structure of the Universe. It evolves $3840^3$ dark matter particles within a cubic volume of side length $1000\,h^{-1}\mathrm{Mpc}$, corresponding to a particle mass resolution of $1.51 \times 10^9\,h^{-1}M_\odot$. MDPL2 adopts a flat $\Lambda$CDM cosmology, consistent with the Planck 2013 results, with parameters ($h$, $\Omega_m$, $\Omega_b$, $n_s$, $\sigma_8$) = (0.6777, 0.307115, 0.048206, 0.96, 0.8228). 

To balance memory constraints with computational efficiency, we down-sample the snapshot at $z=0$ by a sampling rate $0.5\%$ of the entire dataset, yielding approximately a subsample of $2.8 \times 10^8$ particles. This reduced subsample preserves key clustering characteristics and provides a suitable testbed to assess the robustness of our 3PCF measurements and the effectiveness of the associated filtering techniques.  

To measure the 3PCF in a galaxy sample - particularly one with sparse sampling — various statistical estimators have been developed to minimise the shot noise. A widely used edge-corrected estimator, proposed by \citet{SzapudiSzalay1998}, is given by
\begin{equation}\label{eq:SSestimator}
\zeta = \frac{(D - R)^3}{R^3} \equiv \frac{\langle(D_1 - R_1)(D_2 - R_2)(D_3 - R_3)\rangle}{\langle R_1 R_2 R_3 \rangle},
\end{equation}
where the triplet averages are taken over all possible triangle configurations inside the full survey volume. Within the framework of multiresolution analysis, $(D - R)$ can be reconstructed from the differences of the SFCs of the number density fields for the galaxy sample (${\epsilon^D_{j{\bf l}}}$) and the corresponding random sample (${\epsilon^R_{j{\bf l}}}$), explicitly written as $\Delta \epsilon_{j{\bf l}} = \epsilon^D_{j{\bf l}} - \epsilon^R_{j{\bf l}}$.   

The conventional estimation of 3PCF based on Eq.(\ref{eq:SSestimator}) relies on the binning scheme for triplet counting, which in turn depends on the chosen parameterisation of triangle configurations. In \citet{Yue2024}, an alternative binning approach was proposed that is independent of specific triangle shapes. This method performs triplet counting by assigning a finite-volume element to each vertex of the triangle. A natural choice is the spherical tophat window function $W_{\text{TH}}(R)$ with filter radius $R$, which produces a filtered number density field defined as $n_{\text{TH}}({\bf x}) = W_{\text{TH}}(R) \circ n({\bf x})$. In this work, we adopt the same binning scheme.  

In addition, since we analyse simulation data within a periodic box, we directly use the contrast field of the number density $\delta = n/\bar{n} - 1$ as input, where $n$ is the local number density and $\bar{n}$ is its spatial mean. In addition, we adopt reduced 3PCF as our statistical indicator, defined by
\begin{equation}\label{eq:Q-factor}
    \begin{aligned}
    &Q(r_{12},r_{23},r_{31})=\frac{\zeta({\bf r}_1,{\bf r}_2,{\bf r}_3)}{\zeta_H({\bf r}_1,{\bf r}_2,{\bf r}_3)} \\
    &\zeta_H = \xi(r_{12})\xi(r_{23})+\xi(r_{23})\xi(r_{31})+\xi(r_{31})\xi(r_{12}),
    \end{aligned}
\end{equation}
where $r_{ij}=|{\bf r}_i-{\bf r}_j|$ is the distance between ${\bf r}_i$ and ${\bf r}_j$. The reduced 3PCF, also referred to as the hierarchical amplitude $Q$, was first introduced to characterise the original version of the hierarchical clustering models \citep{Peebles1975, Groth_Peebles1977, Peebles1980LSS}. 

To support flexible selection of basis functions, the {\tt PyHermes} toolkit provides an interface to the external wavelet library {\tt PyWavelets}\footnote{\url{https://pywavelets.readthedocs.io/en/latest/}}, an open-source Python package for performing wavelet transforms. Throughout our 3PCF analysis, we adopt Daubechies-4 scaling functions (e.g., \citealt{daubechies1992ten,Fang1998Book}), which have been widely used in cosmological applications (e.g., \citealt{Feng2000,yang2001a,yang2001b,yang2002,Yang2003}). The dilation scale is set to $J = 9$ or $J = 8$, which correspond to spatial resolutions of $1000 / 2^9 = 1.95 h^{-1}\mathrm{Mpc}$ and $3.91 h^{-1}\mathrm{Mpc}$, respectively, for the MDPL2 simulation. In addition, to ensure computational feasibility in practical multipole calculations, the cubic term $(D - R)^3$ in the numerator must be truncated at a specified maximum multipole order. In our numerical tests, this cutoff is set to $l_{\text{max}} = 14$. 

\subsection{Multipole Components: Comparison with the Binning-Corrected Tree-Level Perturbation Theory}

\begin{figure*}
    \centering
    \includegraphics[width=\textwidth]{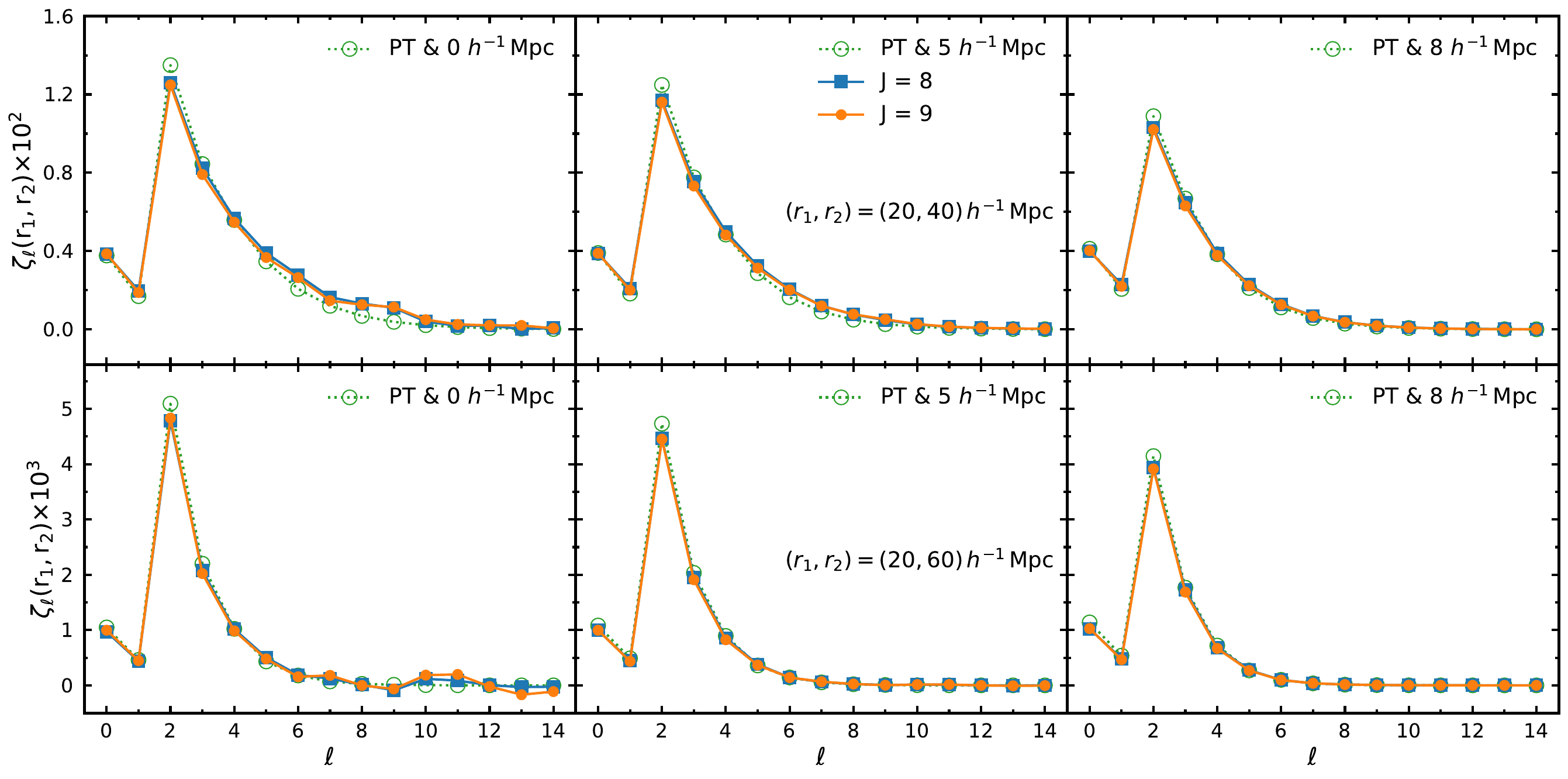}
    \vspace{-0.5cm}
    \caption{The multipole moments for the triangle configurations of $(r_1, r_2) = (20, 40) h^{-1}\mathrm{Mpc}$ (top) and $(20, 60) h^{-1}\mathrm{Mpc}$ (bottom) are shown as a function of the multipole order $L$, measured from the MDPL2 simulation sample. We compare results obtained under different binning conditions: no binning (left, $R = 0 h^{-1}\mathrm{Mpc}$) and with binning using spherical tophat windows of $R = 5 h^{-1}\mathrm{Mpc}$ (middle) and $R = 8 h^{-1}\mathrm{Mpc}$ (right). Additionally, we show measurements at two spatial resolutions, $J = 8$ (square markers with solid lines) and $J = 9$ (circle markers with solid lines). For comparison, the corresponding theoretical predictions from the tree-level PT incorporating binning corrections are also plotted.}
    \label{fig:multipole.vs.PT}
\end{figure*}

Figure~\ref{fig:multipole.vs.PT} presents the multipole moments $\zeta_l(r_1, r_2)$ of the 3PCF for the configurations $(r_1, r_2) = (20, 40) h^{-1}\mathrm{Mpc}$ (top panel) and $(20, 60) h^{-1}\mathrm{Mpc}$ (bottom panel), measured from the MDPL2 simulation and compared with the predictions from the tree-level perturbation theory (PT). For each configuration, we examine two dilation scales, $J=8$ and $J=9$, to investigate how spatial resolution affects convergence. Additionally, an identical tophat window function is applied at each vertex of the triangle. As shown in Fig.~\ref{fig:multipole.vs.PT}, the multipole moment reaches its maximum at the quadrupole $l=2$ and declines rapidly with increasing multipole order $l$. Since both configurations are in the weakly nonlinear regime, tree-level PT works well and is basically consistent with the measurements. 

We further investigated the impact of smoothing on the 3PCF multipole measurements. In Fig.~\ref{fig:multipole.vs.PT}, we compare the results obtained without filtering (left panel, corresponding to a smoothing scale of $R = 0\,h^{-1}\mathrm{Mpc}$) with smoothing scales of $R = 5 h^{-1}\mathrm{Mpc}$ (middle panel) and $R = 8 h^{-1}\mathrm{Mpc}$ (right panel). As expected, increasing the filter scale not only suppresses the amplitude of the multipoles but also smooths out non-linear features on small scales. This trend is clearly illustrated in the figure: the noticeable deviations from tree-level PT predictions in the unsmoothed case become progressively less significant as the smoothing scale increases. The effect is particularly subtle in the weaker non-linear configuration of $(r_1, r_2) = (20, 60) h^{-1}\mathrm{Mpc}$, where the differences are almost indistinguishable.   

Regarding the effect of spatial resolution, Fig.~\ref{fig:multipole.vs.PT} shows only minor differences between the results obtained with $J = 8$ and $J = 9$. This indicates that even the coarser resolution of $J = 8$, corresponding to a spatial scale of $3.91 h^{-1}\mathrm{Mpc}$, is sufficiently fine to resolve the clustering features relevant to the 3PCF at the given triangle configurations. Notably, this resolution remains small compared to the triangle side lengths. As will be discussed later, spatial resolution is a key factor that influences the computational cost of 3PCF measurements. Therefore, selecting an appropriate resolution scale - balancing accuracy and efficiency - is essential for practical analyses. 

\subsection{3PCF: Comparison with Monto Carlo Estimation}

\begin{figure*}
    \centering
    \includegraphics[width=\textwidth]{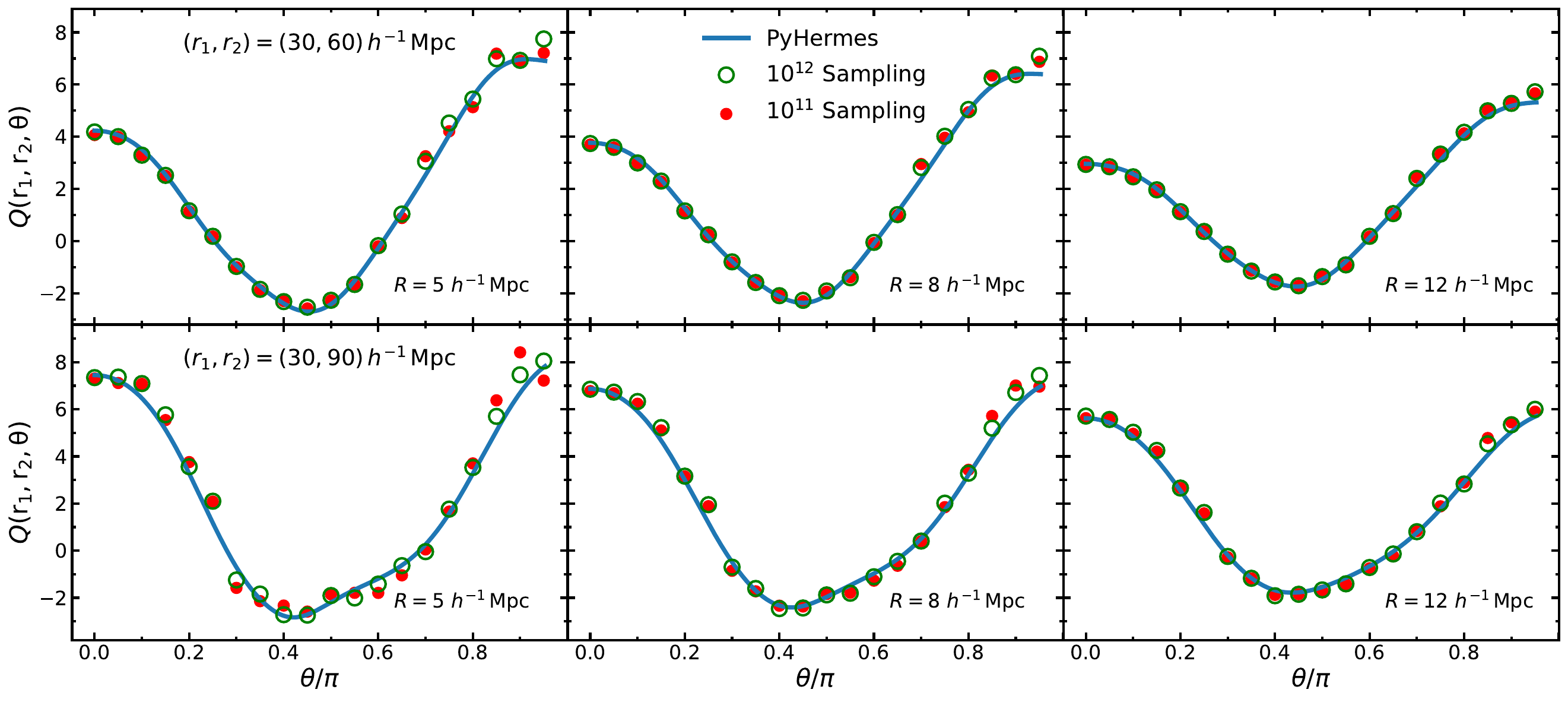}
    \vspace{-0.5cm}
    \caption{The reduced 3PCF in the triangle configurations of $(r_1,r_2)=(30,60)h^{-1}\text{Mpc}$ (upper panel) and $(30,90)h^{-1}\text{Mpc}$ (lower panel) measured in the MDPL2 simulation sample. We compare the results obtained from {\it in situ} multipole method (solid line) with that from the Monte Carlo method with sampling number of $10^{11}$ (solid circle) and $10^{12}$ (open circle). Meanwhile, we apply the binning scheme of spherical top hat with the radius of $R=5h^{-1}\text{Mpc}$ (left), $R=8h^{-1}\text{Mpc}$ (middle), and $R=12h^{-1}\text{Mpc}$ (right).}
    \label{fig:3pcf_hermes_MC}
\end{figure*}

In practice, conventional 3PCF estimation is essentially a Monte Carlo integration: It involves randomly selecting $N_{\text{trans}}$ positions to serve as primary vertices of triangles and applying $N_{\text{rot}}$ random rotations around each primary vertex. This results in a total of $N_{\text{trans}} N_{\text{rot}}$ sampled triangles used in the estimation, which determine the convergence of the result. We mentioned here that some convergence tests are given in the Appendix of \cite{Yue2024}.     

Figure~\ref{fig:3pcf_hermes_MC} compares the 3PCF results obtained using the {\it in situ} multipole method with those from the conventional Monte Carlo experiment. For the latter, we adopt primary sampling points $N_{\text{trans}} = 10^8$ and set the number of random rotations to $N_{\text{rot}} = 10^3$ and $10^4$, resulting in a total of $10^{11}$ and $10^{12}$ triangle configurations sampled, respectively, as indicated in the figure. However, the multipole method represents the 3PCF in terms of its multipole moments. For triangle configurations $(r_1, r_2) = (30, 60)h^{-1}\mathrm{Mpc}$ and $(30, 90)h^{-1}\mathrm{Mpc}$, the multipole amplitudes decay rapidly with increasing $l$. Thus, we safely truncate the multipole expansion at $l = 7$. The analysis is performed on a dilation scale of $J = 8$, corresponding to a grid grid of $256^3$ meshes. 

It is important to emphasise that the multipole method yields a convergent estimation of the 3PCF multipole moments. The complete angular dependence of the 3PCF, i.e., its variation with the angle between the two triangle sides, can be accurately reconstructed as long as the multipole expansion is truncated at a sufficiently large but reasonable maximum order $l$. However, the convergence behaviour of the Monte Carlo method is more sensitive to specific geometric parameters. In particular, two characteristic length scales strongly influence the convergence: the triangle side lengths and the binning size. Longer triangle sides tend to reduce the convergence rate due to the sparsity of contributing triplets, whereas larger bin sizes help accelerate convergence by smearing out the shot noise on small scales. Thus, in the Monte Carlo approach, a careful balance must be struck between angular resolution and computational efficiency. 

Figure~\ref{fig:3pcf_hermes_MC} clearly illustrates the influence of the two side length scales on the accuracy of the 3PCF estimation. The upper and lower rows correspond to triangle configurations with side lengths $(r_1, r_2) = (30, 60) h^{-1}\mathrm{Mpc}$ and $(30, 90) h^{-1}\mathrm{Mpc}$, respectively. As shown, the 3PCF measurements for the longer side configuration (bottom row) exhibit more noticeable fluctuations around the stable reference results produced by the {\tt PyHermes} multipole method with cutoff $l_{\text{max}}=7$, compared to those for the shorter side configuration (top row). This increased jitter reflects the slower convergence of the Monte Carlo method for larger triangle sizes. Similar phenomena have been observed in \cite{Yue2024}. Furthermore, as the smoothing scale $R$ increases from left to right, the level of noise is substantially reduced, demonstrating the regularising effect of smoothing on the Monte Carlo estimation.

\subsection{3PCF: Isosceles Triangle and Nonlinear Effect}

\begin{figure*}
    \centering
    \includegraphics[width=\textwidth]{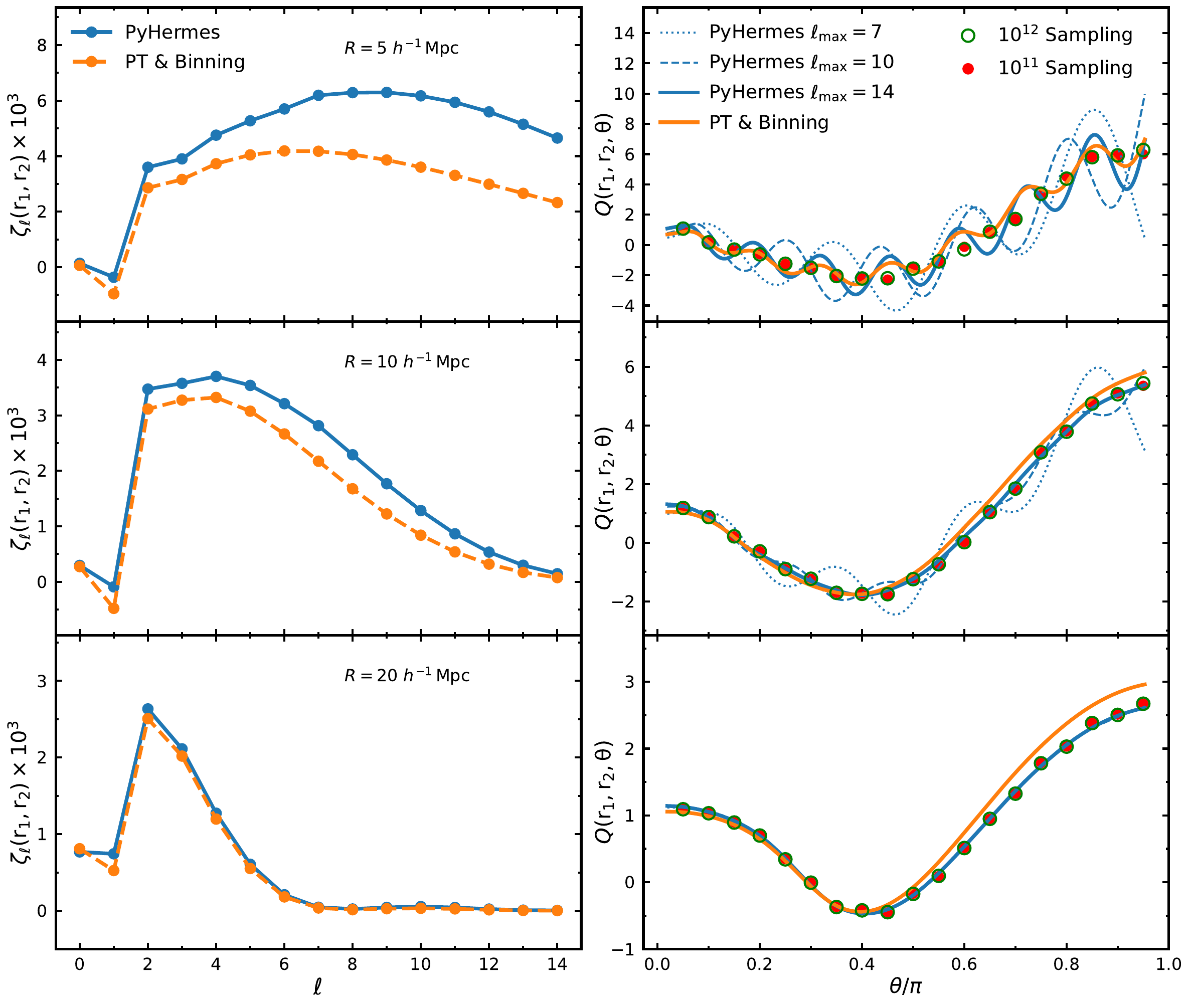}
    \vspace{-0.5cm}
    \caption{Comparison of the 3PCF in an isosceles triangle with equal sides $r_1 = r_2 = 40,h^{-1}\text{Mpc}$ between the MDPL2 simulation and tree-level PT. The spherical tophat binning function is applied with radii of $R = 5 h^{-1}\text{Mpc}$ (top), $R = 10 h^{-1}\text{Mpc}$ (middle), and $R = 20 h^{-1}\text{Mpc}$ (bottom). Left panels: multipole components of the 3PCF measured using {\tt PyHermes}, shown as a function of multipole order up to $l = 14$. Right panels: angle dependence of the reduced 3PCF reconstructed from the multipole components in the corresponding left panels. Reconstructions are made using different truncation orders $l_{\text{max}} = 7$ (dot line), 10 (dash line), and 14 (solid line). To assess convergence, we also show Monte Carlo results with the sampling number of $10^{11}$ (open circles) and $10^{12}$ (solid circles). For comparison, the 3PCF reconstructed from multipoles calculated in the tree-level PT is also shown, in which, we adopt the maximum multipole order $l_{\text{max}}=14$.}   
    \label{fig:isotriangle}
\end{figure*}

We measure the multipole moments of the 3PCF in a special configuration: isosceles triangles with two equal sides, $r_1 = r_2$. This configuration is particularly interesting because for a fixed isosceles side length, the scale of the third side becomes very small when the open angle between the equal sides is sufficiently narrow. Consequently, the third side can probe deeply into the non-linear regime of structure formation. In this regime, the effects of nonlinear gravitational clustering become significant, rendering predictions from conventional tree-level PT unreliable. Figure~\ref{fig:isotriangle} illustrates the multipole moments and the 3PCF constructed with varying truncations for the isosceles triangles with $r_1 = r_2 = 40\text{h}^{-1}\,\text{Mpc}$.

A straightforward way to mitigate the impact of nonlinear effects is to increase the binning width at each vertex of the triangle, effectively enlarging the filtering scale to smooth out small-scale nonlinearities. To examine the influence of binning, we present the measured multipole moments of the 3PCF using various filter scales. For comparison, theoretical predictions based on tree-level PT that incorporate binning effects \citep{Yue2024} are also shown in Figure~\ref{fig:isotriangle}. The left panels display the multipole moments as a function of the multipole order, while the right panels present the constructed 3PCF as a function of the open angle between the equal sides, with different truncations $l_{\mathrm{max}}=7,10,14$ applied in the multipole order. 

Figure~\ref{fig:isotriangle} highlights the significant discrepancies between the measured 3PCF and tree-level PT predictions at the smoothing radii of $R = 5$ and $10h^{-1}\mathrm{Mpc}$ (top left and middle panels), primarily due to nonlinear gravitational clustering. In particular, at $R = 5h^{-1}\mathrm{Mpc}$, the multipole amplitudes exhibit a slow variation with increasing $l$: the values rise gradually from the monopole ($l=0$) through the dipole and quadrupole, peaking around $l = 8$. This behaviour reflects the strong influence of nonlinear structures on small scales. As the smoothing radius increases to $R = 10 h^{-1}\mathrm{Mpc}$, the deviations from the PT predictions become notably smaller. The multipole amplitudes decay more rapidly with $l$, approaching zero near $l = 14$. When the filter radius is further increased to $R = 15 h^{-1}\mathrm{Mpc}$ (bottom row), the measured 3PCF shows good agreement with tree-level PT, demonstrating that larger smoothing scales effectively suppress small-scale nonlinearities and bring the system into the regime where linear theory becomes a valid approximation.  

Since our analysis only includes multipole moments up to $l_{\mathrm{max}} = 14$, we cannot fully assess the behaviour of higher-order moments ($l > 14$), particularly in cases where multipole amplitudes vary slowly with $l$, such as for the small smoothing scale of $R = 5 h^{-1}\mathrm{Mpc}$. To investigate the angular (shape) dependence of the 3PCF and examine the convergence at different truncated orders of multipoles, we reconstruct the reduced 3PCF $Q$ by adding the measured multipoles to the truncations $l_{\mathrm{max}} = 7,10,14$, respectively, as shown in the right panel of Figure~\ref{fig:isotriangle}. Meanwhile, we also plot the theoretical prediction of the tree-level PT with $l_{\mathrm{max}}=14$ and the Monte Carlo results. The latter can be taken as the reference value of the full real 3PCF. To maximise reliability, we record the results at sampling $10^{11}$ and $10^{12}$.  Remarkably, on a smoothing scale of $R = 5 h^{-1}\mathrm{Mpc}$, the reduced 3PCFs reconstructed from the summed multipoles with different truncation orders exhibit pronounced oscillatory behaviour. Similar oscillations are also observed in the theoretical predictions from tree-level PT. Given the fact that individual multipoles vary slowly with $l$, as shown in the left panel, it is evident that none of the reconstructed reduced 3PCF fully converges to the true result, even summing up to $l_{\text{max}}=14$, compared with the reference values from the Monte Carlo estimation. Nevertheless, these oscillations occur around the true values and follow a consistent global trend, suggesting that the essential shape information is still well captured in the angular domain.   

When the smoothing scale is increased to $R = 10 h^{-1}\mathrm{Mpc}$, the oscillations in the measured reduced 3PCF become significantly weaker than those on the smaller smoothing scales. This suppression becomes more pronounced with increasing multipole truncation order. At $l_{\text{max}} = 14$, the reconstructed reduced 3PCF shows good agreement with the Monte Carlo results. On the largest smoothing scale, $R = 20 h^{-1}\mathrm{Mpc}$, we observe excellent consistency between the multipole reconstruction and the true result obtained from the high-precision Monte Carlo calculation.

However, the discrepancy between the measurements and the tree-level PT is still visible in the rising regime of the reduced 3PCF beyond an angle $\theta_{\text{min}}$ at which $Q$ reaches a minimum value. Obviously, the discrepancy increases with the angle at $\theta > \theta_{\text{min}}$. In this triangle configuration, the minimum is reached around $\theta_{\text{min}}=0.4\pi$, corresponding to the opposite side length of $\sim 50h^{-1}\text{Mpc}$. The nonlinearity on large scales arises from the mode-mode coupling. The smoothing due to binning window functions tends to smooth out nonlinearities over scale ranges up to a few smoothing radius; thus, only on even larger scales does nonlinearity become more significant. With the enlarged smoothing radius, the spatial domain affected by the smoothing effect continues to expand, and nonlinearities can only be preserved at larger scales. 



\subsection{Performance Tests}

\begin{figure*}
    \centering
    \includegraphics[width=12.0cm]{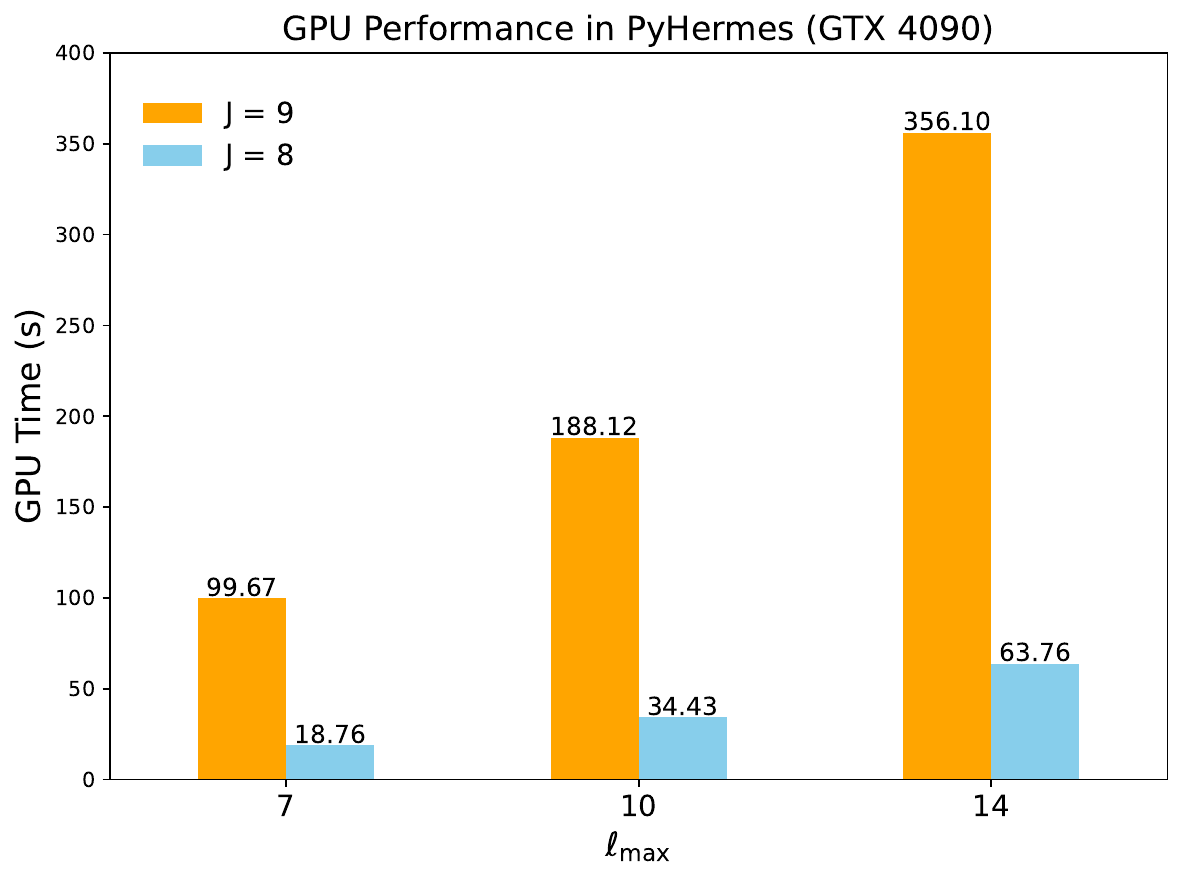}
   \vspace{0.cm}
    \caption{GPU performance test for multipole moment measurements: comparing GPU clock time in units of second for different maximum order of multipoles $l_{\text{max}}=7,10,14$ and two spatial resolutions $J=8$ and $J=9$.}
    \label{fig:benchmark}
\end{figure*} 

The {\it in situ} multipole algorithm has been fully integrated into the {\tt PyHermes} toolkit, which is implemented in Python with support for Message Passing Interface (MPI) parallelisation and GPU acceleration. Since the algorithm reconstructs the density field using a set of basis functions, its computational complexity is independent of the number of objects $N_p$ while ignoring the linear scaling. Instead, the overall computational cost is primarily determined by three integer parameters: 
\begin{itemize}
\item Genus of the father wavelet $D_g$, which controls the linear size of the compact support of the wavelet $D_{\mathrm{sup}}=D_g-1$ and thus affects the local computation volume in summation;
\item Dilation scale $J$ of the scaling function in the multiresolution analysis, which determines the number of grids, $N_g = 2^{3J}$, and thus sets the spatial resolution limit of the reconstructed field;
\item Maximum multipole order $l_{\max}$, which defines the truncation level in the multipole expansion and governs the accuracy of the shape dependence of the 3PCF recovery.
\end{itemize}
According to the workflow illustrated in Fig.~\ref{fig:3PCF-pipline}, the dominant computational costs in the {\it in situ} grid-based multipole algorithm arise from repeated fast Fourier transforms (FFTs) and multidimensional summations. Given the algorithmic parameters discussed above, the FFT operations scale approximately as $\mathcal{O}(l_{\mathrm{max}}^2 N_g \log N_g)$. The multidimensional summation, on the other hand, scales as $\mathcal{O}(D_{\mathrm{sup}}^6 N_g)$. 

In the runtime benchmark, we adopt the Daubechies father wavelet with genus $D_g=4$ and explore dilation scales of $J = 8$ and $J = 9$ to assess the impact of spatial resolution. The multipole expansion is truncated at a maximum order of $l_{\text{max}} = 7,10,14$.  

Runtime performance tests were performed on a high-performance compute server featuring dual {\it AMD EPYC 9754} CPUs (each with 128 cores) and 1TB of shared memory. GPU acceleration was provided by an {\it NVIDIA GeForce RTX 4090} graphics card, equipped with 16,384 CUDA cores and 24GB of {\it GDDR6X} memory. The runtime code was compiled using the NVIDIA CUDA Toolkit version 12.4.

The benchmark results are presented in the histogram shown in Figure~\ref{fig:benchmark}. Recall that for a given $l_{\text{max}}$, the final output of the algorithm consists of a total of $l_{\text{max}} + 1$ multipole moments, representing a compressed form of the 3PCF. As illustrated in the figure, the GPU time exhibits a scaling behaviour approximately proportional to $l_{\text{max}}^2$, consistent with theoretical expectations. For example, in the low-resolution case with $J = 8$, the computation time for $l_{\text{max}} = 7$ is 18.8 seconds, while for $l_{\text{max}} = 14$, it increases to 63.8 seconds by roughly a factor of four, in agreement with the scaling $(14/7)^2$ . When $l_{\text{max}}$ is fixed, the spatial resolution scales with the size of the grid as $N_g = 2^{3J}$, which implies an eight-fold increase in computational workload when moving from $J = 8$ to $J = 9$. However, as shown in Figure~\ref{fig:benchmark}, the actual increase in GPU time is somewhat less than this theoretical prediction. A detailed analysis of task decomposition reveals that this is largely due to nearly constant I/O overhead, which becomes a non-negligible fraction of total runtime at various resolutions.  

\section{Discussion and Concluding Remarks}

This paper presents a highly efficient and scalable algorithm for measuring the multipole moments of the three-point correlation function (3PCF), adaptive to process large datasets from current and upcoming cosmological surveys. The algorithm has been integrated into the {\tt PyHermes} toolkit for cosmic statistics (Ju et al. in preparation). Fundamentally, our approach is parallel in spirit to the multipole-based 3PCF technique developed by \citet{Slepian2015, Slepian2016, Slepian2018}, in which they implemented a fast algorithm with computational complexity $\mathcal{O}(N_p^2)$ for a sample of $N_p$ particles and further accelerated to $\mathcal{O}(N_g \log N_g)$ using FFT on a grid of $N_g$ cells. 

At its core, our {\it in situ} multipole algorithm builds upon a series of foundational concepts developed in \citet{Feng2007} and \citet{Yue2024}, which fundamentally distinguish it from previous approaches. The first key step involves reconstructing the density field from a point process using the framework of multiresolution analysis. Within this framework, a discrete particle distribution—initially represented by singular Dirac delta functions—is approximated by an expansion over compactly supported basis functions at adjustable dilation scales. This reconstruction effectively reduces the aliasing effects commonly introduced by conventional mass assignment/interpolation schemes, such as the cloud-in-cell method (e.g. \citealt{hockney2021computer}), which are typically required for FFT-based algorithms (e.g., \citealt{Slepian2016}). It should be noted that power spectrum estimators that are almost aliasing-free have also been developed within this multiresolution framework \citep{Yang2003, Cui2008}.

Another essential concept underlying our approach is the {\it in situ} perspective on the correlation function \citep{Yue2024}, which leads to recognition of an equivalence between binning, counts in cells, and filtering operations in statistical measurements. This viewpoint offers practical flexibility in choosing window functions to process the density field, encompassing operations such as binning and smoothing. Notably, binned pair counts can, in principle, be performed over arbitrarily shaped geometric volumes, independent of the parameterisation of triangle configurations used in 3PCF. In practice, rather than conventional radial binning, we employ the triple-sphere scheme for triplet counting. As demonstrated in the numerical tests in Section~\ref{NumericalTest}, using spheres with variable radii allows us to control nonlinear clustering and shot noise on small scales. Moreover, enabled by the density field reconstruction, our algorithm can incorporate smooth, non-sharp-edged binning functions tailored to specific scientific objectives. Crucially, binning effects can naturally be incorporated into theoretical predictions \citep{Yue2024}. 

Furthermore, according to the {\it in situ} representation of the 3PCF, the binned counts at the two secondary vertices can be expressed in terms of translated fields centred at the primary vertex, with each translation implemented via a window function. In practical measurements, the 3PCF is computed by statistically averaging all possible triplet counts in a given triangle configuration. Since statistical degrees of freedom arise from translational and rotational invariance, the main algorithmic challenge lies in how to efficiently average over both types of spatial operations. In our approach, as discussed in Section~\ref{subsec:3PCF}, rotational averaging is performed explicitly on the two-fold translation window functions in wavenumber space, as described in Eq.~(\ref{eq:double-e-expansion}), which naturally leads to the multipole expansion technique for the 3PCF. 

For translational invariance, the corresponding average is reduced to a volume integral. Conventionally, this integration is evaluated either by summing over all triangle configurations centred on each particle or via Monte Carlo methods based on random sampling. Alternatively, by reconstructing the density field using a complete set of orthogonal basis functions, the volume integral can be computed analytically. In the case of the 2PCF, which involves a product of two density fields, the orthogonality of the basis functions simplifies the integral to a scalar product of two data vectors. Extending this to the 3PCF, we encounter a trilinear form of the 3-fold {\it in situ} density field. As shown in Section~\ref{subsec:3PCF}, by introducing the connection coefficients $\Gamma$ associated with the basis functions, the trilinear form can be reduced to a bilinear one, resulting in a double summation.

We note that the {\it in situ} multipole framework has been extended to NPCF for arbitrary $N$ \citep{Philcox2022, PhilcoxEtal2022}, with an implementation available in the {\it ENCORE} package. Building on this, our future work aims to extend our multipole-based algorithm to the NPCF, utilising the isotropic basis functions developed by \citet{Cahn2023}, and combining with the {\it in situ} formalism of the correlation function and the density field reconstruction scheme based on multiresolution analysis. Furthermore, incorporating anisotropic effects into the multipole algorithm will be essential for accurately modelling nonlinear clustering in redshift space \citep{Slepian2018}, and we plan to revisit this issue in future work. 

As an alternative to measuring the 3PCF as a function of the opening angle in fixed triangle configurations (parameterized by two side lengths), the multipole moments of the 3PCF offer an efficient form of data compression. They characterise the angular dependence of galaxy clustering and enhance sensitivity to nonlinearities, thereby potentially placing tighter constraints on the nonlinear growth history of large-scale structures. According to our performance tests, the high computational efficiency of the multipole algorithm enables exploration of nonlinear effects across a large configuration space. This, in turn, allows for improved constraints on background cosmology, galaxy biasing models, and primordial non-Gaussianities. With the advent of ongoing and forthcoming galaxy surveys—such as Euclid \citep{Laureijs2011}, CSST \citep{Zhan2011}, the Dark Energy Spectroscopic Instrument (DESI) \citep{DESI2016}, the Nancy Grace Roman Space Telescope \citep{Akeson2019}, and the Vera C. Rubin Observatory \citep{Ivezic2019}—which will map the Universe with tens of millions to billions of galaxies, our multipole-based 3PCF algorithm provides a practical and scalable statistical tool for cosmological applications.           

 
\section*{Acknowledgements}

FLL is supported by the National Key R\&D Program of China through grant 2020YFC2201400 and the NSFC Key Program through the grants 11733010 and 11333008. ZH is supported by the NFSC general program (grant no. 12073088), the National key R\&D Program of China (grant no. 2020YFC2201600), and the National SKA Program of China no. 2020SKA0110402. For the purpose of open access, the author has applied a Creative Commons Attribution (CC BY) license to any Author Accepted Manuscript version arising from this submission.

The CosmoSim database used in this paper is a service by the Leibniz-Institute for Astrophysics Potsdam (AIP). The MultiDark database was developed in cooperation with the Spanish MultiDark Consolider Project CSD2009-00064. The authors gratefully acknowledge the Gauss Centre for Supercomputing e.V. (\url{www.gauss-centre.eu}) and the Partnership for Advanced Supercomputing in Europe (PRACE, \url{www.prace-ri.eu}) for funding the MultiDark simulation project by providing computing time on the GCS Supercomputer SuperMUC at the Leibniz Supercomputing Centre (LRZ, \url{www.lrz.de}). 

\section*{Data Availability}

All simulation data analysed in this paper are available through the MultiDark database website \url{https://www.cosmosim.org}. The authors agree to make data supporting the results or analyses presented in their paper available upon reasonable request. The {\tt PyHermes} toolkit will be publicly available on GitHub together with the release of the relevant code paper soon (Ju et al., in preparation). An early test version of {\tt PyHermes} is also available from the corresponding author upon special request. 



\bibliographystyle{mnras}
\bibliography{OptAlgorithm3PCF} 




\appendix

\section{Proof of spherical harmonics expansion of dual plane waves}\label{sec:proofs} 

Applying the spherical harmonics expansion Eq.(\ref{eq:plane-sph-harmonics}) of plane wave on both $e^{i\mathbf{k}_1\cdot \mathbf{r}_1}$ and $e^{i\mathbf{k}_2\cdot \mathbf{r}_2} $, we obtain
\begin{equation}
\left\langle e^{i\mathbf{k}_1\cdot \mathbf{r}_1} e^{i\mathbf{k}_2\cdot \mathbf{r}_2} \right\rangle = 16\pi^2 \sum_{\ell_1=0}^{\infty}\sum_{\ell_2=0}^{\infty}i^{\ell_1+\ell_2} j_{\ell_1}(k_1r_1)j_{\ell_2}(k_2r_2)\sum_{m_1=-\ell_1}^{\ell_1}\sum_{m_2=-\ell_2}^{\ell_2} Y_{\ell_1 m_1}\left(\hat{\mathbf{k}}_1\right)Y_{\ell_2 m_2}^*\left(\hat{\mathbf{k}}_2\right)\left\langle Y_{\ell_1 m_1}^*\left(\hat{\mathbf{r}}_1\right) Y_{\ell_2 m_2}\left(\hat{\mathbf{r}}_2\right)\right\rangle . \label{eq:eik_ave}
\end{equation}
The average $\left\langle Y_{\ell_1 m_1}^*\left(\hat{\mathbf{r}}_1\right) Y_{\ell_2 m_2}\left(\hat{\mathbf{r}}_2\right)\right\rangle$ can be equivalently computed by fixing $\mathbf{r}_1$ and $\mathbf{r}_2$ and randomly rotating the coordinate system. We choose the initial spherical coordinate system so that $\hat{\mathbf{r}}_1$ is the north direction, and denote the spherical harmonics in this particular coordinate system as $\tilde{Y}_{\ell m}$. It then follows that
\begin{equation}
    \tilde{Y}_{\ell m}(\hat{\mathbf{r}}_1) = \sqrt{\frac{2\ell+1}{4\pi}}\delta_{m0}, \label{eq:Ylm_i1}
\end{equation} 
and
\begin{equation}
    \tilde{Y}_{\ell 0}(\hat{\mathbf{r}}_2) = \sqrt{\frac{2\ell+1}{4\pi}}P_\ell(\hat{\mathbf{r}}_1\cdot \hat{\mathbf{r}}_2). \label{eq:Ylm_i2}
\end{equation} 
When the coordinate system is rotated by Euler angles $(\alpha,\beta, \gamma)$ in the $z$-$y$-$z$ convention, the spherical harmonics in the rotated coordinate systems are 
\begin{equation}
Y_{\ell m}(\mathbf{n}) = \sum_{m^\prime=-\ell}^\ell D^\ell_{mm^\prime}(\alpha,\beta,\gamma) \tilde{Y}_{\ell m^\prime}(\mathbf{n}), 
\end{equation}
where $D^\ell_{mm^\prime}(\alpha,\beta,\gamma)$ is the Wigner D-matrix.  Thus, we have
\begin{equation}
\left\langle Y_{\ell_1 m_1}^*\left(\hat{\mathbf{r}}_1\right) Y_{\ell_2 m_2}\left(\hat{\mathbf{r}}_2\right)\right\rangle = \sum_{m_1^\prime=-\ell_1}^{\ell_1} \sum_{m_2^\prime=-\ell_2}^{\ell_2} 
 \left\langle {D^{\ell_1}_{m_1m_1^\prime}}^*(\alpha,\beta,\gamma)D^{\ell_2}_{m_2m_2^\prime}(\alpha,\beta,\gamma) \right\rangle \tilde{Y}_{\ell_1 m_1^\prime}^*(\hat{\mathbf{r}}_1)  \tilde{Y}_{\ell_2 m_2^\prime}(\hat{\mathbf{r}}_2). \label{eq:coortrans}
 \end{equation}
Applying Schur orthogonality relations on the SO(3) group yields \citep{brink1993angular}
\begin{equation}
     \left\langle {D^{\ell_1}_{m_1m_1^\prime}}^*(\alpha,\beta,\gamma)D^{\ell_2}_{m_2m_2^\prime}(\alpha,\beta,\gamma) \right\rangle = \frac{1}{8\pi^2}\int_0^{2\pi}\mathrm{d}\alpha\int_0^{\pi}\sin\beta~ \mathrm{d}\beta\int_0^{2\pi}{D^{\ell_1}_{m_1m_1^\prime}}^*(\alpha,\beta,\gamma)D^{\ell_2}_{m_2m_2^\prime}(\alpha,\beta,\gamma)~\mathrm{d}\gamma  =  \frac{1}{2\ell_1+1}\delta_{\ell_1\ell_2}\delta_{m_1m_2}\delta_{m_1^\prime m_2^\prime}. \label{eq:orth}
\end{equation}
Putting Eqs.~(\ref{eq:Ylm_i1},\ref{eq:Ylm_i2},\ref{eq:coortrans},\ref{eq:orth}) together, we have
\begin{equation}
    \left\langle Y_{\ell_1 m_1}^*\left(\hat{\mathbf{r}}_1\right) Y_{\ell_2 m_2}\left(\hat{\mathbf{r}}_2\right)\right\rangle = \frac{1}{4\pi} P_{\ell_1}(\hat{\mathbf{r}}_1\cdot \hat{\mathbf{r}}_2)\delta_{\ell_1\ell_2}\delta_{m_1m_2}. \label{eq:Ylm_ave}
\end{equation}
We now substitute Eq.~\eqref{eq:Ylm_ave} into Eq.~\eqref{eq:eik_ave} and obtain
\begin{equation}\label{eq:double-e-expansion-sep}
    \left\langle e^{i\mathbf{k}_1\cdot \mathbf{r}_1} e^{i\mathbf{k}_2\cdot \mathbf{r}_2} \right\rangle = 4\pi \sum_{\ell=0}^{\infty}(-1)^\ell  j_{\ell}(k_1r_1)j_{\ell}(k_2r_2) P_\ell(\hat{\mathbf{r}}_1\cdot \hat{\mathbf{r}}_2)\sum_{m=-\ell}^{\ell} Y_{\ell m}\left(\hat{\mathbf{k}}_1\right)Y_{\ell m}^*\left(\hat{\mathbf{k}}_2\right). 
\end{equation}
Substituting the addition theorem 
\begin{equation}
\sum_{m=-\ell}^{m=\ell} Y_{\ell m}(\hat{\mathbf{k}}_1)^* Y_{\ell m}(\hat{\mathbf{k}}_2)=\frac{2 \ell+1}{4 \pi}P_{\ell}(\hat{\mathbf{k}}_1 \cdot \hat{\mathbf{k}}_2)
\end{equation}
into the above equation yields the desired Eq.~\eqref{eq:double-e-expansion}.


\bsp	
\label{lastpage}
\end{document}